\documentstyle[12pt]{article}
\def\bar{\overline}
\def\ZZ{\mbox{\rm Z}\hskip-4pt \mbox{\rm Z}}

\def\NN{\mbox{\rm I}\hskip-2pt \mbox{\rm N}}
\def\CC{\mbox{\rm C}\hskip-6pt\mbox{l} \;}
\def\id{\mbox{\rm 1}\hskip-3pt \mbox{\rm l}}

\def\LD{{\cal L}}
\def\HD{{\cal H}}

\def\dd{\mbox{d}}
\newcommand{\bg}{\begin{equation}}
\newcommand{\eg}{\end{equation}}
\def\su21{$SU(2\, \vert \, 1)$}
\begin{document}
\phantom{.} \hfill MZ-TH/93-26\\
\phantom{.} \hfill October 1993
\begin{center}
{\large MODELS OF ELECTROWEAK INTERACTIONS IN\\[6pt]
NON-COMMUTATIVE GEOMETRY: A COMPARISON$^{\ast )}$}
\end{center}
\vspace{2cm}
\begin{center}
N.A. Papadopoulos, J. Plass, and F. Scheck$^1$\\
Institut f\H ur Physik\\Johannes Gutenberg-Universit\H at\\
D-55099 Mainz (Germany)\\
\small
e-mail: SCHECK@VIPMZA.UNI-MAINZ.DE
\normalsize
\end{center}
\vspace{2cm}

\begin{abstract}
Alain Connes' construction of the standard model is based on a generalized
Dirac-Yukawa operator and the K-cycle $(\HD ,D)$, with $\HD$ a fermionic
Hilbert space. If this construction is reformulated at the level of the
differential algebra then a direct comparison with the alternative approach
by the Marseille-Mainz group becomes possible. We do this for the case of
the toy model based on the structure group $U(1)\times U(1)$ and for the
$SU(2)\times U(1)$ of electroweak interactions. Connes' results are
recovered without the somewhat disturbing $\gamma_{5}$-factors in the
fermion mass terms and Yukawa couplings. We discuss both constructions in
the same framework and, in particular, pinpoint the origin of the
difference in the Higgs potential obtained by them.
\end{abstract}

\vfill
{\small $^{\ast)}$ Work supported in part by PROCOPE project
Mainz University and CPT Marseille-Luminy\\
$^{1}$ Also at CERN, CH-1211 Geneva 23}
\newpage
{\bf 1.}
The minimal standard model of strong and
electroweak interactions is in perfect agreement
with experiment (see e.g. \cite{Okun}). As experiment allows for only small
deviations from its predictions there is not much freedom for extended or
alternative descriptions of
the fundamental interactions. Recently, in the framework of non-commutative
geometry, several constructions were proposed which either are very
close to or yield
exactly the standard model \cite{CONN}-\cite{HPS}. The non-commutative
geometry approach extends and generalizes the framework of Yang-Mills
theories but, in the end, yields the same Yang-Mills-Higgs Lagrangian. A
most noteworthy virtue of this new approach is the progress in
understanding better some of the {\it qualitative\/} features
of the standard model. For instance, the Higgs fields appear naturally
as part of a (super)connection. Spontaneous symmetry breaking
receives a new geometric interpretation \cite{HPS,CHPS}. This should be
compared to the usual, more {\it quantitative\/}, phenomenology of the Higgs
sector which concentrates on determining its parameters from
experiment \cite{Kniel}.

The aim of this note is to compare the construction of the standard model
proposed by Connes \cite{CONN} and elaborated further by Kastler
\cite{KAST}, to the approach developed by the Marseille group and the Mainz
group \cite{CEV}-\cite{CHPS}. In the sequel, we
shall refer to the former as construction (I), and to the latter as
construction (II), for the sake of brevity.

Connes' construction (I) is formulated in terms of abstract algebraic
objects which generalize, in the sense of non-commutative geometry, the
notion of differential forms and, as a result, the notion of connection
(gauge potential). The generalized differential forms are realized within a
certain fermionic representation space and by means of a (generalized)
Dirac-Yukawa operator. After a lengthy calculation an action is obtained
which is very close to, albeit not identical with, the one of the
standard model. When taken at face value and at the classical level, the
Lagrangian obtained in (I) fixes some of the parameter ratios of the model
\cite{KSCH}. The first formulation of (I) contained auxiliary, adynamical
fields which had to be eliminated by minimization and which made the
comparison with (II) difficult. Since then Connes proposed a modified
construction of the underlying differential algebra \cite{CONN1} which
renders such a comparison much easier.

The Marseille-Mainz construction (II) rests on less sophisticated
mathematics. It is formulated in the $\ZZ_{2}$-graded space of
matrix-valued differential forms, equipped with a multiplication law
inherited, in a straightforward manner, from the tensor product of matrix
multiplication and the wedge product for differential forms. It makes use
of a generalized differential which is composed of the usual exterior
(Cartan) derivative and a matrix derivative. Construction (II) leads to a
Lagrangian which coincides exactly with the Lagrangian of the standard
model, including spontaneous symmetry breaking and in the correct
``shifted'' phase of the neutral Higgs field. Again,
some ratios of the model's parameters seem to be fixed but they can be
modified even at the classical level \cite{SUSY}. Furthermore,
there is nothing to protect these ratios under quantization \cite{AGM}.

With the new formulation of (I) referred to above a detailed comparison of
the two approaches is now within reach because Connes' modified
differential algebra and ours are very similar. In what follows we
work out this comparison in the simplest case of one generation of
leptons. The results of this comparison are interesting and, to some
extent, surprising.
We start with a short summary of (II). We then present the main
steps which lead to (I) in a framework which is very close to the one of
(II). In this way we achieve not only a direct and transparent comparison
but we are enabled to perform calculations within Connes' framework (I)
which are considerably simpler than in the original work \cite{KAST}. In
order to facilitate the understanding we present our detailed calculations
only for the case of the structure group $U(1)\times U(1)$, considered in
\cite{CEV}. We make this restriction because in the framework of (I)
calculations are simplified considerably as compared to
the more realistic case of the structure group $SU(2)\times U(1)$.
The conclusions are exactly the same. We show that the $\gamma_{5}$ factor in
the fermionic mass terms, not present in the standard model and also not
present in (II), can in fact be avoided in the approach (I). Finally, we
comment on the question of parameter ratios. We show that the
situation is very similar in either construction, i.e. that the conclusions
reached in (II) also apply to (I).

\vspace{12pt}
{\bf 2.} In summarizing (II) we closely follow \cite{CEV}.
The basic mathematical
object is the $\ZZ_{2}$-graded algebra $\Omega_{M}^{\ast}(X)$ of
matrix-valued forms as obtained from the skew tensor product of the matrix
algebra $M(n,\CC )$ and the algebra $\Lambda^{\ast}(X)$ of differential
forms over space-time $X$. The matrix algebra is is taken to be
$\ZZ_{2}$-graded, and $\Lambda^{\ast}$ carries its $\NN$-grading. So we
have
\[ \Omega_{M}^{\ast}(X)=M(n,\CC )\widehat{\otimes}\Lambda^{\ast}(X) \, .\]
Here we take $n=2$. In this case the $\ZZ_{2}$-(matrix) grading
distinguishes the diagonal part, which is even, and the off-diagonal part,
which is odd, and can be defined with the grading automorphism
$ \Gamma = \mbox{diag}(1,-1)$.
Homogeneous elements of $\Omega_{M}^{\ast}$ are written  as
$a\otimes \alpha$, with $a$ a matrix and $\alpha$ a differential form. In a
simplified notation the grade of this element is
\[ \partial (a\otimes \alpha )=\partial a+\partial\alpha
   \, (\mbox{mod }2)\, , \]
and the multiplication law reads
\[ (a\otimes\alpha )\bullet (b\otimes\beta )=(-)^{\partial\alpha
\cdot\partial b}a\cdot b\otimes \alpha\wedge\beta \, .\]
The generalized differential is given by
\[ \dd (a\otimes\alpha )=(\dd_{M}a)\otimes\alpha +(-)^{\partial a}
a\otimes \dd_{C}\alpha \, , \]
with $\dd_{C}$ denoting the usual (Cartan) exterior derivative in
$\Lambda^{\ast}$, and $\dd_{M}$ the matrix derivative in $M(2,\CC )$
defined by its action on the even and odd parts $a_{0}$ and $a_{1}$ of
$a$, respectively, \cite{HPS}
\[ \dd_{M}(a)=[\eta ,a_{0}]+i\{\eta ,a_{1}\} \, , \]
with $\eta$ denoting the odd element of $M(2,\CC )$
\bg \eta = i \left(\begin{array}{cc} 0 & c \\ \bar{c} & 0 \end{array}
\right) \, .\eg
In an obvious notation the structure of the algebra $\Omega_{M}^{\ast}$ is
summarized by writing symbolically
\bg \left( \Omega_{M}^{\ast}(X),\bullet ,\dd \right)_{\ZZ_{2}} \, . \eg
The generalized potential (superconnection) $\cal A$ is an element of
$\Omega_{M}^{1}(X)$ and reads explicitely\footnote{We choose conventions
such that $\cal A$ and $\cal F$ are antihermitean, cf. \cite{CHPS}}
\bg {\cal A}=i \left( \begin{array}{cc}
A & c\Phi /\mu \\ \bar{c}\bar{\Phi}/\mu & B
\end{array}\right) \eg
with $A=A_{\mu}\dd x^{\mu},B=B_{\mu}\dd x^{\mu}$, $\Phi$ a scalar field,
$c$ a constant and
$\mu$ a parameter of dimension mass . The field strength (supercurvature)
is obtained, e.g., form the structure equation
\[ {\cal F}=\dd {\cal A} +{\cal A}\bullet {\cal A} \, . \]
Without loss of generality we may set $c=1$ and $\mu =1$, (the
dimensionless $c$ can be absorbed in the mass parameter $\mu$
while the latter sets the mass scale).
The explicit form of $\cal F$ is then
\bg {\cal F}=i \left( \begin{array}{cc}
\dd_{C}A-(\Phi+\bar{\Phi}+\Phi\bar{\Phi}) &
-\dd_{C}\Phi -i(A-B)(\Phi+1) \\
-\dd_{C}\bar{\Phi} +i(A-B)(\bar{\Phi}+1) &
\dd_{C}B-(\Phi+\bar{\Phi}+\Phi\bar{\Phi})
\end{array}\right) \, . \eg
The Lagrangian is calculated from
$\LD =-\mbox{tr }({\cal F}^{\dagger}{\cal F})$ and takes the form
\bg \LD = -\frac{1}{4}F_{\mu\nu}^{A}F^{A\mu\nu}
          -\frac{1}{4}F_{\mu\nu}^{B}F^{B\mu\nu}+
2{\cal D}\bar{\Phi}{\cal D}\Phi - V(\Phi) \, \eg
with ${\cal D}\Phi=D\Phi +i(A-B)$, $D\Phi =\dd_{C}\Phi +i(A\Phi -\Phi B)$,
and $V(\Phi)=2(\Phi +\bar{\Phi}+\Phi\bar{\Phi})^{2}$.
It is important to notice that the Higgs potential stems from the diagonal
part of $\cal F$, while the mass term of the boson field $Z=A-B$ originates
from the off-diagonal. It is evident that (5) is the exact analogue of the
standard model Lagrangian. In particular, through the structure equation
it includes spontaneous symmetry breaking and places the Higgs-like
field $\Phi$ in the right ``shifted'' phase. As discussed in detail in
\cite{CES,HPS,CHPS} an analogous result is obtained
in the case of the structure
group $SU(2)\times U(1)$ and of the graded Lie algebra $su(2\vert 1)$.

\vspace{12pt}
{\bf 3.} We now describe approach (I) in a formulation which follows closely
the discussion of (II) given above. Regarding the specific question of
constructing the standard model in noncommutative geometry, we thus
provide a third derivation of (I), after Connes' original work
\cite{CONN,CONN1} and Kastler's detailed account of (I) \cite{KAST},
which is considerably simpler and more transparent than the original
formulation.

Denote by $\cal F$ the space of
complex functions on spacetime $X$ and, for the case of $U(1)\times U(1)$,
write
\[ {\cal M}=\left(\begin{array}{cc} \CC & 0 \\ 0 & \CC \end{array}
\right) \, . \]
Connes' approach (I) is based on the algebra
${\cal A}={\cal M}\otimes {\cal F}$ and the corresponding universal
differential envelope $(\Omega^{\ast}({\cal A}),\delta )$ that is
generated by the formal elements (``words'')
$A_{0}\delta A_{1}\ldots \delta A_{n}\in \Omega^{n}({\cal A})$ and
the operator $\delta$ obeying the Leibniz rule
$\delta (AB)=(\delta A)B+A(\delta B)$. This algebra is realized by means of
a K-cycle (Dirac-Kasparov cycle) $({\cal H},D)$ over $\cal A$, where
$\cal H$ is a Hilbert space and $D$ a Dirac-Yukawa operator, and a
representation $\pi$ of $\Omega^{\ast}({\cal A})$ on that Hilbert space.
The Dirac-Yukawa operator has the form
$D=i\gamma^{\mu}\partial_{\mu}+D_{M}$ where $D_{M}=\mu\eta$, with $\eta$ as
given in the construction of (II) above, cf. eq. (1). $D_{M}$ may be
understood to be a fermionic mass matrix. Note that such an interpretation
was not made in (II) because it is unnecessary in that framework. The
representation $\pi$ of the universal envelope on the space $\LD (\HD )$
of bounded linear operators over $\cal H$ is given by
\begin{eqnarray}
\pi :\,\Omega^{\ast}({\cal A}) & \longrightarrow & \LD (\HD )\nonumber\\
A_{0}\delta A_{1}\ldots\delta A_{n} & \longrightarrow &
A_{0}[D,A_{1}]\ldots [D,A_{n}]\; .
\end{eqnarray}
In the original version of Connes' construction the gauge potential and the
field strength were taken to be elements of $\pi (\Omega^{\ast}({\cal A}))$.
This led to the appearence of auxiliary or adynamic fields (fields without
kinetic energy) in the Lagrangian which had to be eliminated by
minimization \cite{KAST,CFF}. At that stage a direct comparison with other
approaches such as (II) was impossible.

In the more recent version of (I) given in \cite{CONN1}
one goes one step further by
considering the space $\Omega_{D}^{\ast}({\cal A})$, obtained from
$\Omega^{\ast}({\cal A})$ by dividing out the ideals
$J^{k}({\cal A})=(K^{k}+\delta K^{k-1})$, where
$K^{k}:=\mbox{ker }\pi \cap \Omega^{k}$, viz.
\[ \Omega_{D}^{k}({\cal A})=\Omega^{k}({\cal A})/J^{k}({\cal A})\, , \]
or, equivalently,
\[\Omega_{D}^{k}({\cal A})=\pi (\Omega^{k}({\cal A}))/
\pi (J^{k}({\cal A}))\]
In contrast to $\pi (\Omega^{\ast}({\cal A}))$ the space $\Omega_{D}^{\ast}$
is an $\NN$-graded differential algebra (like the universal object
$\Omega^{\ast}$). Therefore, $\Omega_{D}^{\ast}({\cal A})$ is the space
which should be compared to the space $\Omega_{M}^{\ast}(X)$ of the
approach (II) discussed in sect. 2 above. The multiplication law is defined
by the ordinary multiplication in $\LD (\HD )$ and by taking the quotient.
We denote it by the symbol $\odot$. The differential, denoted by $\delta$,
is given by commutation with the Dirac-Yukawa operator and by taking the
quotient as above. In obvious analogy to (2) we may summarize the structure
of this algebra as follows
\bg \left( \Omega_{D}^{\ast}({\cal A}),\odot ,\delta \right)_{\NN} \, .\eg

The explicit construction of the space $\Omega_{D}^{\ast}({\cal A})$ in the
most general case, to the best of our knowledge, has not been given in the
literature. There is, however, important progress in this direction that
will be published elsewhere \cite{KPPW}. For the example of
$U(1)\times U(1)$ (and likewise for the case of $SU(2)\times U(1)$) the
explicit calculation can be performed and the results for
$\Omega_{D}^{\ast}$, $\odot$, and $\delta$ can be given in simple terms.
This is what we set out to do next.

For the purposes of physics we need to know only the spaces
$\pi (\Omega^{k})$ for $k=$ 0, 1, and 2. Thus we have to determine
the projected ideals $\pi (J^{k})$ for these three values of $k$. They are
found to be, respectively,
\bg \pi (J^{0})=\{ 0\}, \quad \pi (J^{1})=\{ 0\}, \quad
\pi (J^{2})={\cal M}_{0}\otimes \Lambda^{0}(X) \, , \eg
so that we have
\bg \Omega_{D}^{0}=\left(\begin{array}{cc}
\Lambda^{0} & 0 \\ 0 & \Lambda^{0}\end{array}\right)=
{\cal M}_{0}\otimes\Lambda^{0}(X)\, , \eg
\bg \Omega_{D}^{1}=\left(\begin{array}{cc}
\Lambda^{1} & 0 \\ 0 & \Lambda^{1}\end{array}\right)+
\left(\begin{array}{cc}
0 & \Lambda^{0}\\\Lambda^{0} & 0\end{array}\right)\equiv
{\cal M}_{0}\otimes \Lambda^{1}(X)+{\cal M}_{1}\otimes \Lambda^{0}(X)
\, , \eg
\begin{eqnarray}
\Omega_{D}^{2} & = & \frac{{\cal M}_{0}\otimes \Lambda^{2}+
{\cal M}_{1}\otimes \Lambda^{1}+{\cal M}_{0}\otimes \Lambda^{0}}
{{\cal M}_{0}\otimes \Lambda^{0}}
\nonumber\\ & \cong & {\cal M}_{0}\otimes\Lambda^{2}(X)+
{\cal M}_{1}\otimes \Lambda^{1}(X)\nonumber\\
 & = & \left(\begin{array}{cc}
\Lambda^{2} & 0 \\ 0 & \Lambda^{2} \end{array}\right) +
\left(\begin{array}{cc}
0 & \Lambda^{1} \\ \Lambda^{1} & 0 \end{array}\right)\, . \end{eqnarray}
The construction for grades 0, 1, and 2 is fairly obvious\footnote{A more
detailed derivation of eq. (11) is given in \cite{KPPW}. Note that for
{\it more\/} than one generation $\Omega_{D}^{2}$ would also contain zero
forms.}.
For arbitrary grade $k\in \NN$, $\Omega_{D}^{k}$ can be shown to be
given by the pattern visible already in (10) and (11), viz.
\[ \Omega_{D}^{k}({\cal A})=\left(\begin{array}{cc}
\Lambda^{k}(X) & 0 \\ 0 & \Lambda^{k}(X)\end{array}\right)+
\left(\begin{array}{cc}
0 & \Lambda^{k-1}(X) \\ \Lambda^{k-1}(X) & 0\end{array}\right)\, . \]
Regarding the multiplication law we derive the example of the product
$\Omega_{D}^{1}\times \Omega_{D}^{1}\rightarrow\Omega_{D}^{2}$. The general
case can easily be guessed from this example. We have, in an obvious
notation,
\begin{eqnarray} \left[\left(\begin{array}{cc}
\Lambda^{1} & 0 \\ 0 & \Lambda^{1}\end{array}\right)+
\left(\begin{array}{cc}
0 & \Lambda^{0} \\ \Lambda^{0} & 0\end{array}\right)\right]
& \odot &
\left[\left(\begin{array}{cc}
\Lambda^{1} & 0 \\ 0 & \Lambda^{1}\end{array}\right)+
\left(\begin{array}{cc}
0 & \Lambda^{0} \\ \Lambda^{0} & 0\end{array}\right)\right]
\nonumber\\
=\left(\begin{array}{cc}
\Lambda^{2} & 0 \\ 0 & \Lambda^{2}\end{array}\right)+
\left(\begin{array}{cc}
0 & \Lambda^{1} \\ \Lambda^{1} & 0\end{array}\right) & &\, .
\end{eqnarray}
In particular, it is important to note that
\[ \left(\begin{array}{cc}
0 & \Lambda^{0}(X) \\ \Lambda^{0}(X) & 0\end{array}\right)\odot
\left(\begin{array}{cc}
0 & \Lambda^{0}(X) \\ \Lambda^{0}(X) & 0\end{array}\right) =
0\in \Omega_{D}^{2} \, . \]
It is easy to work out the action of the differential $\delta$ on
$\Omega_{D}^{\ast}$. Again we give only the example of grade $k=1$,
$\delta :\Omega_{D}^{1} \rightarrow \Omega_{D}^{2}$ :
\begin{eqnarray}
\delta\left[\left(\begin{array}{cc}
\Lambda^{1} & 0 \\ 0 & \Lambda^{1}\end{array}\right)+
\left(\begin{array}{cc}
0 & \Lambda^{0} \\ \Lambda^{0} & 0\end{array}\right)\right] & = &
\dd_{C}\left(\begin{array}{cc} \Lambda^{1} & 0 \\ 0 & \Lambda^{1}
\end{array}\right) -
\dd_{C}\left(\begin{array}{cc} 0 & \Lambda^{0} \\ \Lambda^{0} & 0
\end{array}\right)\nonumber\\
 & &
 + \left[\eta ,\left(\begin{array}{cc}\Lambda^{1} & 0 \\
0 & \Lambda^{1}\end{array}\right)\right]+
i\left\{ \eta ,\left(\begin{array}{cc} 0 & \Lambda^{0}\\ \Lambda^{0} & 0
\end{array}\right)\right\} \, .
\end{eqnarray}
Note that for the same reasons as in the multiplication law the
anticommutator on the right-hand side vanishes,
\[ \left\{ \eta ,\left( \begin{array}{cc}
0 & \Lambda^{0} \\ \Lambda^{0} & 0 \end{array}\right)\right\}
= 0 \in \Omega_{D}^{2} \, . \]
Thus, we have at our disposal an explicit construction of the space (7).

The construction of the generalized potential and field strength proceeds
along the same lines as for (II), cf. sect. 2, keeping track of the
modified multiplication and differential. As the spaces of grade 1 are the
same in both frameworks, $\Omega_{M}^{1}(X)=\Omega_{D}^{1}({\cal A})$, the
gauge potential (3) is the same. The field strength given by
\[ {\cal F} := \delta {\cal A} + {\cal A}\odot {\cal A} \, , \]
however, is different. A straightforward calculation using the
multiplication rule and the differential given above leads to the result
\bg {\cal F}=i\left( \begin{array}{cc}
\dd_{C}A & -\dd_{C}\Phi -i(A-B)(\Phi+1) \\
-\dd_{C}\bar{\Phi} +i(A-B)(\bar{\Phi}+1) &
\dd_{C}B \end{array}\right) \, . \eg
The most noticeable difference to eq. (4) is that the Higgs potential has
disappeared from eq. (14). Indeed, the corresponding Lagrangian is given by
\bg \LD = -\frac{1}{4}F_{\mu\nu}^{A}F^{A\mu\nu}
          -\frac{1}{4}F_{\mu\nu}^{B}F^{B\mu\nu}+
2{\cal D}\bar{\Phi}{\cal D}\Phi \, . \eg
This is a consequence of the fact that we considered one generation of
fermions (leptons) only. Thus, with this assumption, (I) leads to a trivial
Higgs potential while (II) yields the correct potential and spontaneous
symmetry breaking in the correct ``shifted'' phase. If one adds one or
more generations then the Higgs potential appears also in (I) provided
the fermion masses are not degenerate (see below).

\vspace{12pt}
{\bf 4.} We conclude by sketching the analogous calculation in the more
realistic case of $SU(2)\times U(1)$. The Marseille-Mainz construction
(II), described in sect. 2 above, comes closest to Connes' result if $\eta$
is chosen as in eq. (1) above with
\[ c= \left(\begin{array}{c} 1 \\ 0 \end{array}\right)\otimes M \]
and $M$ a fermionic mass matrix, possibly containing more than one
generation. At the same time, the ansatz for the (super)connection
$\cal A$ is enlarged by tensorizing $\Phi$ with $M$, $\bar{\Phi}$ with
$M^{\dagger}$.
Repeating the calculation of the field strength $\cal F$ whose
details are found in \cite{CES,CHPS} one finds the expressions given in
\cite{CHPS,FS} with the following modifications. Denote for a moment by
${\cal F}^{(0)}$ the field strength obtained with the choice
$c= (1,0)^{T}$ in eq. (1). In the diagonal elements ${\cal F}_{11}^{0}$
and ${\cal F}_{22}^{0}$ of $\cal F$ the terms
in the Higgs fields are tensorized with $MM^{\dagger}$ and $M^{\dagger}M$,
respectively, while the off-diagonal terms are tensorized with $M$ and
$M^{\dagger}$, respectively,
\[ {\cal F}_{12}={\cal F}_{12}^{(0)}\otimes M \qquad
    {\cal F}_{21}={\cal F}_{21}^{(0)}\otimes M^{\dagger} \, . \]
As a consequence the kinetic energy of the Higgs field in the Lagrangian
(5) is multiplied by $\mbox{tr}(MM^{\dagger})$, while the Higgs potential is
multiplied by $\mbox{tr}(MM^{\dagger})^{2}$.

Regarding Connes' construction (I), the essential difference with (II) lies
in the diagonal elements of $\cal F$.  Before taking the quotient, they
contain terms of the type
$F_{\mu\nu}\gamma^{\mu}\gamma^{\nu}$ and terms of the form
$(\Phi+\bar{\Phi}+\bar{\Phi}\Phi)MM^{\dagger}\id$. The division by the
ideal $J^{2}$, in essence, leads to replacement of $MM^{\dagger}$ by
\bg (MM^{\dagger})_{\perp}=MM^{\dagger}-\frac{1}{n}\mbox{tr}(MM^{\dagger})
\, , \eg
where $n$ is the dimension of the matrix $(MM^{\dagger})$. As a result, the
Higgs potential is proportional to $\mbox{tr}(MM^{\dagger})_{\perp}^{2}$
and vanishes whenever $(MM^{\dagger})$ is proportional to the unit matrix,
i.e. when the masses of equally charged fermions are degenerate
\cite{VaBo}. This is the main difference in the two approaches.

\vspace{12pt}
In conclusion we have developed a simplified, somewhat more algebraic
construction of Connes' algebra which allows for a direct comparison with
the Marseille-Mainz approach. Unlike Connes' approach we did not start from
the generalized Dirac-Yukawa operator. Thereby we avoid the somewhat
disturbing factors $\gamma_{5}$ in the off-diagonal elements of that
operator which lead to fermionic mass matrices and to Yukawa couplings
which are not those of the standard model. By deriving both (I) and
(II) in the same algebraic spirit we exhibit more clearly the similarities
and localize the differences in the underlying algebras (2) and (7). Our
explicit calculation of the parts with grade 0, 1, and 2 of these algebras
shows that while in (II) the Higgs potential is independent of fermionic
masses altogether, the approach (I) yields a non-vanishing potential only
if these masses are not degenerate.

\vspace{24pt}
\small
\begin{center} {\bf Acknowledgement}\end{center}
We thank W. Kalau and J.M. Warzecha for discussions. F.S. acknowledges
gratefully the kind hospitality extended to him by CERN's theory group
during his sabbatical term 1993.
\normalsize
\newpage

\end{document}